# In Cloud, Do MTC or HTC Service Providers Benefit from the Economies of Scale?

Lei Wang, Jianfeng Zhan
Chinese Academy of Sciences
{wl, jfzhan}@ncic.ac.cn

Weisong Shi
Wayne State University
weisong@wayne.edu

Yi Liang, Lin Yuan
Beijing University of Technology;Chinese Academy of Sciences
yliang@bjut.edu.cn;yuanlin@ncic.ac.cn

## ABSTRACT

Cloud computing, which is advocated as an economic platform for daily computing, has become a hot topic for both industrial and academic communities in the last couple of years. The basic idea behind cloud computing is that resource providers, which own the cloud platform, offer elastic resources to end users. In this paper, we intend to answer one key question to the success of cloud computing: in *cloud*, do many task computing (*MTC*) or high throughput computing (*HTC*) service providers, which offer the corresponding computing service to end users, benefit from the economies of scale? To the best of our knowledge, no previous work designs and implements the enabling system to consolidate MTC and HTC workloads on the cloud platform and no one answers the above question. Our research contributions are three-fold: first, we propose an innovative usage model, called ***dynamic service provision (DSP) model***, for MTC or HTC service providers. In the DSP model, the resource provider provides the service of creating and managing runtime environments for MTC or HTC service providers, and consolidates heterogeneous MTC or HTC workloads on the cloud platform; second, based on the DSP model, we design and implement ***Dawningcloud***, which provides automatic management for heterogeneous workloads; third, a comprehensive evaluation of Dawningcloud has been performed in an emulatation experiment. We found that for typical workloads, in comparison with the previous two cloud solutions, Dawningcloud saves the resource consumption maximally by 46.4% (HTC) and 74.9% (MTC) for the service providers, and saves the total resource consumption maximally by 29.7% for the resource provider. At the same time, comparing with the traditional solution that provides MTC or HTC services with dedicated systems, Dawningcloud is more cost-effective. To this end, we conclude that for typical MTC and HTC workloads, on the cloud platform, MTC and HTC service providers and the resource service provider can benefit from the economies of scale.

**Categories and Subject Descriptors:** D.4.7 [**Operating Systems**] Organization and Design - *distributed systems*

## 1. INTRODUCTION

Many-task computing (*MTC*) can deliver much large numbers of computing resources over short period of time to accomplish many computational tasks [1], and high throughput computing (*HTC*) can deliver large amounts of processing capacity over long period of time [1]. Traditionally, many small or medium organizations tend to *purchase* and *build* dedicated cluster systems (*DCS*) to provide computing services for MTC or HTC applications. We call this usage model *the DCS model* and the corresponding system *the DCS system*. The DCS model prevails in MTC and HTC communities, and the organization owns a small or medium-scale cluster system and deploys the specific runtime environment for MTC or HTC workloads. With the *full control,* the administrators of the DCS systems manage the affiliated user accounts and configure the related management policies for the specific runtime environments, such as scheduling or resource management policies*.* However, there are also two shortcomings of the DCS model: first, the total cost of ownership (*TCO*) is high, which includes the cost of power, manpower, equipment depreciation, etc; second, for peak loads, the DCS systems can not provide enough resources, while lots of resources are idle for normal loads.

Recently, as resource providers or infrastructure providers [2], several pioneer computing companies are advocating infrastructure as a service [2]. For example, as a resource provider, Amazon [3] has provided elastic computing cloud (EC2) service to offer outsourced resources to end users at the granularity of XEN [4] virtual machines. A new term cloud is used to describe this new computing paradigm. Cloud is a large pool of easily usable and accessible virtualized resources, which can be dynamically reconfigured and typically exploited by a pay-per-use model [6]. Though a cloud system may imply geographically distributed systems [8], in this paper, when we refer to a cloud platform, it indicates a centralized cluster system. In 2006, we have coined a new term *Industrial Information Grid (IIG)* [24], similar to the concept of *open federated cloud computing* [8], to describe the system that exclusively own geographically distributed resources for Web service application.

In this paper, we want to focus the key issues to the success of cloud computing: *for small or medium organizations, can we consolidate their MTC and HTC workloads on a large cloud platform? And on the cloud platform, do MTC or HTC service providers benefit from the economies of scale?*

Previous efforts [5] [7] have validated the possibility of running HPC applications on cloud platforms. However, to the best of our knowledge, previous efforts fail to resolve the above issues in several ways. First, there are two proposed usage models for cloud computing in *MTC* or *HTC* communities. Deelman *et al.* [10] propose that each staff of an organization (end use) directly leases virtual machine resources from EC2 in a specified period for running applications, and we call this usage model the *direct resource provision (DRP)* model and the corresponding system the *DRP system*. In DRP, each end user rents resources from the resource provider directly. Our experiment results show that the DRP system will lead to high *peak* resources consumption, which raises challenge for the capacity planning of system. Evangelinos *et al.* [5] propose that the organization as a whole rents resources with the fixed size from EC2 to create a virtual cluster system that is deployed with the queuing system, like OpenPBS, for HTC workloads. We call this usage model the *static service provision (SSP) model* and the corresponding system the *SSP system*. In SSP, a service provider as a whole leases the resources with fixed size from the resource provider, deploys a PBS-like queuing system and provides job-execution services for end users. Our experiment results show that for typical workloads, the SSP system leads to high resource consumption because of its static resource management policy.

Second, previous efforts fail to propose the enabling system with the autonomic management mechanism to facilitate the resource provider to consolidate MTC and HTC workloads: EC2 [3] directly provides resources to end users, and relies upon end user's manual management of resources; EC2 extended services: RightScale [5] provides automated cloud computing management systems that helps you create and deploy *only Web service applications* that run on the EC2 platform; Irwin *et al.* [20] [13] propose a prototype of service oriented architecture for resource providers and consumers to negotiate access to resources over time. However, no previous effort proposes the autonomic management system to consolidate MTC and HTC workloads.

Third, no previous work answers this key question: do MTC or HTC service providers benefit from the economies of scale? Armbrust *et al.* [2] in theory show the workloads of *Web service applications* can benefit from the economies of scale of cloud computing systems. Our previous work, Phoenixcloud [12] [21], shows the consolidation of *Web service applications* and *parallel batch jobs* can decrease the total resource consumption from the perspective of the resource provider.

On the Dawning 5000 cluster system, which is ranked as top 10 of Top 500 super computers in November, 2008 [14], we design and implement an innovative system *DawningCloud*. With the enabling DawningCloud system, the organization does not need to own a DCS system, and instead the resource provider is responsible for managing and monitoring a cluster-based cloud platform, creating the specific runtime environment for a MTC or HTC service provider, and dynamically provisioning resources to runtime environments; the administrator of an organization, as the proxy of a service provider, manages its runtime environment with the *full controls* and provides MTC or HTC service to its end users; as end users, the staffs in the organization use the web portal of the runtime environment to submit and manage their MTC or HTC applications.

The contributions of our paper can be concluded as follows:

First, in cloud, we propose a new usage model, called *dynamic service provision (DSP) model*. Similar to DCS and SSP models, service providers in the DSP model can *fully control* their runtime environments; and unlike the SSP and DCS models, service providers can dynamically resize the resources according to the workload status.

Second, based on the DSP model, we design and implement DawningCloud. DawningCloud creates runtime environments on the demand for MTC or HTC service providers, and automatically provisions resources to runtime environments.

Thirdly, we conduct a comprehensive evaluation of the DSP model with the enabling system: DawningCloud. Our experiments show that using DawningCloud, MTC and HTC service providers benefit from the economies of scale. In our experiments, for typical MTC workload: Montage workflow, and typical HTC workload traces: NASA iPSC trace and SDSC BLUE trace, in comparison with the DRP system, DawningCloud saves the resource consumption maximally by 46.4% (HTC) and 74.9% (MTC) for the service providers, and saves the total resource consumption by 29.7% for the resource provider; in comparison with the SSP and DCS systems, DawningCloud saves the resource consumption maximally by 32.5% (HTC) for the service providers, and saves the total resource consumption by 29.7% for the resource provider, moreover, DawningCloud is more cost-effective than the DCS system through the cost analysis of a real case. This result implies that DawningCloud can achieve the economies of scale for the resource provider, and MTC or HTC service providers can benefit from the economies of scale in cloud.

The organization for the rest of the paper is as follows: Section 2 describes the proposed DSP model; Section 3 gives out the design and implementation of DawningCloud; Section 4 systematically compare Danwingcloud with the other system: SSP, DCS and DRP; Section 5 summarizes the related work; Section 6 draws the conclusion and discusses the future work.

## 2. THE DSP MODEL

In this section, we proposed the *DSP model*: first, we describe the roles in a cloud platform; second, we introduce the details of the DSP model; third, we conclude the distinguished differences of the DSP model from the DRP, SSP and DCS models.

### 2.1 Three Players in a Cloud

We propose three roles in the cloud platform: *resource provider, service providers and end users*.

*Resource provider:* the resource provider owns a cloud platform, and offers outsourced resources. The typical example is Amazon.

Different from EC2 of which the resource provider directly offers resources to ends user, we propose another role, *service provider*, which acts as the proxy of an organization. The service providers lease the resources from the resource provider and provide computing service to its end uses. The staffs in the organization play the role of the *end users*. In a typical cloud platform, there are only one resource provider, several service providers and their affiliated end users.

### 2.2 DSP Details

In the DSP model, the resource provider is responsible for managing the cloud platform, creating the specific runtime environment for a MTC or HTC service provider, and provisioning resources to runtime environments. In the rest of this section, we introduce the usage pattern of the DSP model.

As shown in Fig.1, the usage pattern is described as follows:

1) A service provider specifies its requirement for runtime environment (*RE*), including types of workloads: MTC or HTC, size of resources, types of operating system, and then *requests* the resource provider for creating the customized RE. In our technical report [21], we have given out a description model for describing the diversities of requirements of different service providers.

2) The resource provider *creates* the RE for the service provider according to its requirement. Section 3.1.3 will report the lifecycle management mechanism.

3) After the RE is created, the service provider can *manage* his RE with the full control, e.g. creating accounts for end users.

4) End users use their accounts to *submit* and *manage* MTC or HTC applications in the RE.

5) When the RE is providing services, according to the current workload, the RE can automatically *negotiate resources* with the proxy of the resource provider to resize resources by leasing more resources or releasing idle resources.

6) If the service provider wants to terminate his computing service, the service provider will inform end users to backup. End users can backup their data to the storage server provided by the resource provider. And then the service provider will *destroy accounts* of the end users' in the RE.

7) The service provider *confirms* the resource provider that the RE is ready for destroying.

8) The resource provider *destroys* the specified RE and withdraws the corresponding resources.

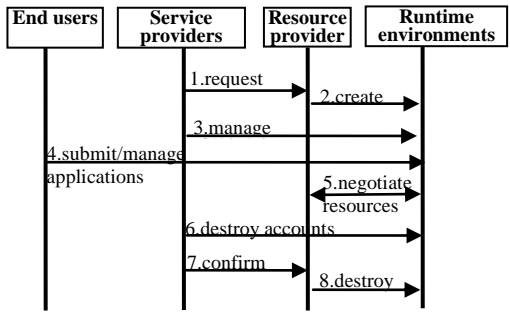

**Figure 1. The usage pattern of DSP.**

## 2.3 The Distinguished Differences of the DSP Model

Table 1 summarizes the differences of DCS, SSP, DRP and DSP usage models.

**Table 1. The comparison of different usage models**

|  | DCS | SSP | DRP | DSP |
|---|---|---|---|---|
| resource property | local | leased | leased | leased |
| runtime environment | stereo-typed | stereo-typed | no offering | created on the demand |
| resources provision for RE | fixed | fixed | manual | flexible |

There are two main distinguished differences of the DSP model from the other models.

First, in the DSP model, on the cloud platform the resource provider can create the specific runtime environments on the demand for MTC service providers or HTC service providers.

This property does not hold true for the DRP model. In the DRP model, there is no runtime environment that automatically manages resources for MTC or HTC workloads, and each end user directly obtains resources from the resource providers. In the SSP model, the service provider is limited in that he requests the resources with the fixed size and deploys a batch queuing system for HTC workloads. In the DCS model, a service provider owns the resources locally, and provides stereotyped MTC or HTC RE.

Second, in the DSP model, the service provider can dynamically resize the provisioned resources. The property does not hold true for the DRP, SSP and DCS models. In the DRP model, each end user manually requests or releases resources from the resource provider. In the SSP model, the organization as a whole obtains the resources with the fixed size from the resource provider. In the DCS model, the service provider purchases and builds a dedicated cluster system with the fixed size.

## 3. Enabling System: DawningCloud

To provide MTC or HTC services, different organizations have different research plans, and their workloads may vary in the same period. We argue that on a cloud platform, the consolidation of different workloads of MTC and HTC may achieve the economies of scale for the resource provider. So, according to the DSP model and our previous Phoenixcloud system [12] [21], we design and implement an enabling system, DawningCloud, for the resource provider to consolidate MTC and HTC workloads.

In this section, we introduce two most important features of DawningCloud: first, how to create a runtime environment on the demand for a MTC or HTC service provider on a cloud platform? Second, we propose an automatic resource management mechanism for coexisting runtime environments of different service providers.

## 3.1 Creating Runtime Environment on the Demand for MTC or HTC Workloads

### 3.1.1 The Requirement Differences of MTC and HTC Runtime Environments

Since there are diversities of MTC workloads [1] and HTC workloads, in this paper, we take a typical MTC workload, Montage workflow [23], and a representative HTC workload, batch jobs, to present the design of MTC and HTC runtime environments. Montage workflows are introduced as a typical MTC workload in the work of Ian Foster [1], and batch jobs are also presented as the representative HTC workloads in the condor project [19]. In the DawningCloud design, we consider three requirement differences of runtime environments between MTC and HTC workloads as follows:

1) The usage scene: the aim of HTC is designed for running parallel/sequential batch jobs; the aim of MTC is designed for running scientific workflows, like Montage workflow [1].

2) The application characteristic: MTC applications [1] can be decomposed to a set of small jobs with dependencies, whose running time is short; while batch jobs in HTC are independent and the running times of jobs are varying.

3) The evaluation metric: HTC service providers concern the job's throughput over long period of time; while MTC

service providers concern the job's throughput over short periods of time.

### 3.1.2 The Layered Architecture of DawningCloud

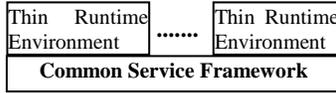

**Figure 2. The framework of DawningCloud.**

As shown in Figure 2, we present a layered architecture for DawningCloud: one is the *common service framework (CSF)* and the other is the *thin runtime environment (TRE)*. The concept of TRE [21] indicates that the common sets of functions for different runtime environments are delegated to the CSF, and a TRE only implements the core functions for the specific workload.

The major functions of the CSF are responsible for managing the lifecycle of TREs, for example creating, destroying TREs, and provisioning resources to TREs in terms of nodes or virtual machines. The main services of the CSF [21] are as follows:

*The resource provision service* is responsible for providing resources to different TREs.

*The lifecycle management service* is responsible for managing the lifecycle of TREs.

*The deployment service* is a collection of services for deploying and booting operating system, the CSF and TREs.

*The virtual machine provision service* is responsible for creating or destroying virtual machine like XEN.

*The agent* is responsible for downloading the required software package, starting or stopping service daemon.

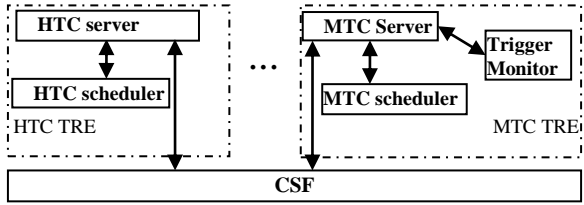

**Figure 3. Coexisting MTC and HTC runtime environments.**

In DawningCloud, we implement two types of TREs: MTC TRE and HTC TRE.

In HTC TRE, we only implement three services: *the HTC scheduler, the HTC server and the HTC web portal*. The HTC scheduler is responsible for scheduling the user's job through scheduling policy. The HTC server is responsible for dealing with users' requests, managing resources, loading jobs. The HTC web portal is the GUI through which end users submit and monitor HTC applications.

In MTC TRE, we implement four services: the MTC scheduler, the MTC server, the trigger monitor and the MTC web portal. The function of the *MTC scheduler* is similar to the HTC scheduler. Different from the HTC server, the *MTC server* needs to parse the workflow description model, which are inputted by users on the MTC web portal, and then submit a set of jobs with dependencies to the MTC scheduler for scheduling. Besides, a new service, named the *trigger monitor*, is responsible for monitoring the trigger condition of workflows, such as the changes of database's record or files, and notifying the changes to the MTC server to drive the running of jobs in different stages of a workflow. The *MTC web portal* is also much more complex than that of HTC, since it needs to provide a visual editing tool for end users to draw different workflows.

Figure 3 shows a typical DawningCloud system, of which a MTC TRE and a HTC TRE reuse the CSF.

### 3.1.3 The Lifecycle Management of TREs

The CSF is responsible for managing the lifecycle of a TRE. We introduce this feature taking a MTC TRE as an example. The lifecycle management of a MTC TRE is as follows:

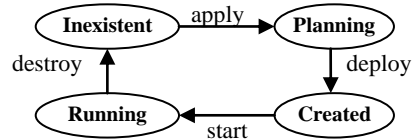

**Figure 4. The lifetime of a TRE.**

1) As shown in Figure 4, the initial state of a MTC runtime environment is *inexistent*. The service provider uses the web portal of the CSF to *apply* for a new MTC TRE. The web portal of the CSF sends the requesting information to the lifecycle management service of the CSF.

2) The lifecycle management service validates the information. If the requesting information is valid, it marks the state of new MTC TRE as *planning*.

3) The lifecycle management service sends *the message of deploying TRE* to agents of the CSF on the related nodes, which requests the deployment service to download the required software package of the MTC TRE. After the new MTC TRE is deployed, the lifecycle management service marks its state as *created*.

4) The lifecycle management service sends the configuration information of the new MTC TRE to the resource provision service of the CSF.

5) The lifecycle management service sends the message to agents to start the components of the new MTC TRE, including the MTC server, the MTC scheduler, the trigger monitor and the MTC web portal. When the MTC server is started, the command parameters will tell it what configuration parameters should be read. Then the lifecycle management service marks the state of the new MTC TRE as *running*.

6) The new MTC TRE begins providing service to end users. End users use the MTC web portal to submit their applications.

7) According to the load status, the MTC server dynamically requests or releases resources from or to the resource provision service.

8) If the service provider uses the web portal of CSF to *destroy* his MTC TRE, the web portal of the CSF sends the destroying information to the lifecycle management service; the lifecycle management service validates the information and destroys the MTC TRE through prompting end users to backup data, stopping the related daemons and offloading the related software packages.

## 3.2 The Automatic Resource Management

### 3.2.1 Dynamic Resource Negotiation Mechanism
We present the dynamic resource negotiation mechanism in DawningCloud as follows:

1) The service provider specifies his requirement for resource management in *the resource management policy, which defines the behavior specification of the server in that the server resizes resource to what an extent according to what criterion.* According to *the resource management policy*, the MTC or HTC server decides whether and to what an extent resizes the resource according to the current workload status, and then sends the requests of obtaining or releasing resources to the resource provision service.

2) The resource provider can specify his requirement for resource provision in *the resource provision policy, which determines when the resource provision service provisions how many resources to different TREs in what priority.* According to *the resource provision policy*, the resource provision service decides to assign or reclaim how many resources to or from the TRE.

3) *The setup policy determines when and how to do the setup work, such as wiping off the operating system or doing nothing.* The resource provision service notifies the negotiation result to the server. At the same time, for each assigned or reclaimed node, *the setup policy* is triggered, and the resource provision service requests the lifecycle management service to do the setup work.

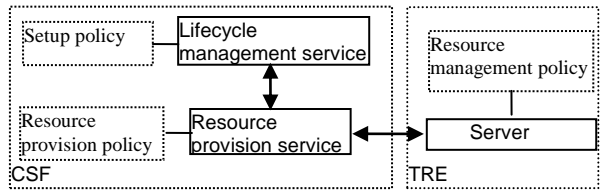

**Figure 5. Dynamic resource negotiation mechanism.**

### 3.2.2 Resource Management and Provision Policies
In this section, we respectively propose resource management and provision policies for MTC and HTC service providers.

#### 3.2.2.1 The Resource Management Policy for HTC
We propose the resource management policy for HTC service provider as follows:

There are two kinds of resources provisioned by the runtime environment: *initial resources* and *dynamic resources.* Once allocated to the TRE, initial resources will not be reclaimed by the resource provision services until the TRE is destroyed. On the contrary, dynamic resources assigned to the TRE may be reclaimed by the resource provision service.

(1) At the startup, the HTC service provider will request *initial resources*.

We define *the ratio of obtaining resources* as the ratio of the accumulated resource demands of all jobs in the queue to the current resources owned by a HTC TRE. When requesting to create a runtime environment, the service provider will set a *threshold ratio of obtaining resources*. For a TRE, when the current ratio of obtaining resources exceeds the threshold ratio, it implies that many jobs need to be queued unless the server can request more resources.

(2) The server of the HTC TRE scans jobs in queue per minute. If *the ratio of obtaining resources* exceeds *the threshold ratio of obtaining resources*, the HTC server will request the dynamic resources with the size of *DR1* as follows:

*DR1=the accumulated resources demand of all jobs in the queue – the current resources owned by the TRE.*

After obtaining enough resource from the resource provision service, the HTC server registers a timer, once per hour, to check idle resources. If there are idle resources with the size equal with or more than the value of DR1, the server will release the resources with the size of the DR1 to the resource provision service.

(3) The HTC server scans jobs in queue per minute. If the ratio of *the resource demand of the present biggest job in the queue* to *the current resources owned by a TRE* is greater than *one* and *the ratio of obtaining resources* does not exceed *the threshold ratio of obtaining resources,* the server will request the dynamic resources with the size of *DR2* as follows:

*DR2= the resources needed by the job with the largest resources demand – the current resources owned by the TRE.*

When the ratio of the resource demand of the present biggest job in the queue to the current resources owned by a TRE is greater than one, it indicates that if the HTC server does not request more resources, the present biggest job may not have enough resources for running.

After obtaining enough resources from the resource provision service, the server registers a timer, once per hour, to check idle resources. If there are idle resources with the size equal with or more than the value of DR2, the server will release the idle resources with the size of the DR2 to the resource provision service.

#### 3.2.2.2 The Resource Management Policy for MTC
The resource management policy of the MTC service provider differs from that of HTC service providers in two aspects. First, the server scans jobs with different intervals. An HTC server scans jobs in queue *per one minute*, while a MTC server scans jobs in queue *per three seconds.* This is because MTC tasks often run over in seconds. Second, when the MTC server calculates the accumulated resource demands of all jobs in queue or the resource demand of the present biggest jobs, each job in queue that constitutes a workflow is calculated. However, for HTC, each independent job in queue is calculated.

#### 3.2.2.3 The Resource Provision Policy for MTC and HTC
We propose a simple resource provision policy for MTC and HTC as: First, the resource provision service provisions enough initial resources to the TRE at the startup of the TRE; Second, when the server of a TRE requests dynamic resources, the resource provision service either assigns enough resources to the server or rejects if the resource provision service has no enough resources; Third, when the server of a TRE releases dynamic resources, the resource provision service will passive reclaim all the released resources.

## 4. PERFORMANCE EVALUATION

In this section, we answer the key question: *do MTC or HTC service providers benefit from the economies of scale*. We choose three workloads from three different organizations. Among them, there are *only one* resource provider, two organizations providing HTC services and one organization providing MTC service. The resource provider respectively chooses DawningCloud, the DRP, SSP and DCS systems to provide computing service. We will compare DawningCloud with the SSP, DRP and DCS system.

### 4.1 Evaluation Method

The period of a typical workload trace is often weeks. To evaluate the system, many key factors have effects on the experiment results, and we need do many times of time-consuming experiments. So we use the emulation method to speedup experiments.

Figure 6, Figure 7 and Figure 8 respectively show the emulated DawningCloud, DRP, SSP and DCS systems.

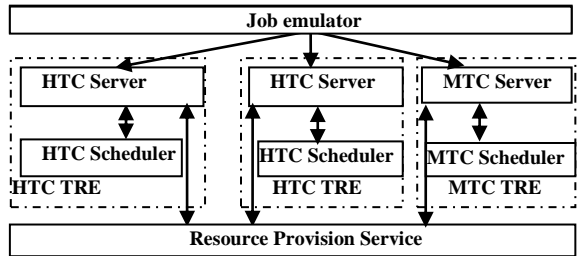

**Figure 6. The emulated DawningCloud.**

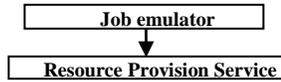

**Figure 7. The emulated DRP system.**

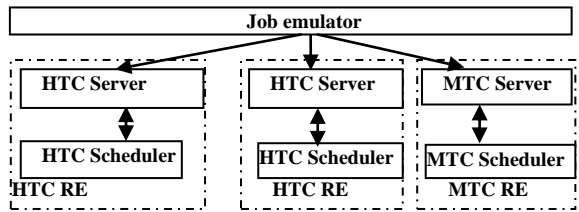

**Figure 8. The emulated SSP and DCS system.**

For all emulated systems, the job emulator is used to emulate the process of submitting jobs. For HTC workload, the job emulator generates jobs by reading the trace file, and then submits jobs. For MTC workload, the job emulator reads the workflow file, generates each job that constitutes each workflow and their dependencies between each job, and then submits jobs according to the dependency constraints. We speed up the submission and completion of jobs by a factor of 100.

In comparison with the real DawningCloud system, our emulated system of DawningCloud for three service providers and one resource provider keeps the resource provision service, two HTC servers, one MTC servers, and three schedulers, while other services are removed, as shown in Figure 6.

As shown in Figure 7, our emulated DRP system only includes the resource provision service and the job emulator, since in the real system, each end user of three organizations directly leases resources from the resource provider, like EC2.

As shown in Figure 8, our emulated SSP and DCS systems include two HTC servers, one MTC server and three schedulers. We remove the resource provision service because the service provider in the SSP model obtains all resources with the fixed size from the resource provider once the runtime environment is created and three service providers in the DCS model owns the fixed resources. So the emulated systems do not need to reflect the interactions between service providers and the resource provider.

### 4.2 Workloads

For MTC, we choose a typical workload, Montage workflow [23], which is an astronomy workflow application, created by NASA/IPAC Infrared Science Archive for gathering multiple input images to create custom mosaics of the sky. The workload generator can be found on the web site [15], and the workload file includes the task name, run time, inputs, outputs and the list of control-flow dependencies of each job. The chosen Montage workload includes 1,000 tasks and the average execution time of tasks is 11.38 seconds.

We choose two typical HTC workload traces from [17]. The utilization rate of all traces in [17] varies from 24.4% to 86.5%. We choose one trace with lower load: the NASA iPSC trace and one trace with higher load: the SDSC BLUE trace. The NASA trace is lower load with 46.6% utilization, while the BLUE trace is higher load with 76.2% utilization. The scales of NASA trace and BLUE trace are respectively 128 and 144 nodes, which are popular in small organizations.

The SDSC BLUE trace is of two weeks from Apr 25 15:00:03 PDT 2000. In the first half of the trace, the job arrived infrequently; in the second half of the trace, the job arrived frequently. The NASA iPSC trace is of two weeks from Fri Oct 01 00:00:03 PDT 1993 and the arrived jobs varied each day.

### 4.3 Evaluation Metrics

We choose *the number of completed jobs* in a certain period [16] to evaluate the performance metric of the HTC service providers; and we choose *tasks per second* [1] to reflect the performance metric of the MTC service providers. For a service provider, we choose the *resource consumption* in terms of *node\*hour* to evaluate the cost. In the DRP system, there is no role of the service provider, so we calculate the accumulated resource consumption of all end users for a workload. For the DCS system, since the service provider owns the resources, we calculate the resource consumption of the service provider as the product of the configuration size of the DCS system and the period of the workload.

For the resource provider, we choose *the total resource consumption* in terms of *node\*hour* to evaluate the economies of scale. In addition, we specially care about *the peak resource consumption* that is a key factor in capacity planning for a resource provider.

### 4.4 Experiment Design and Configuration

Since the workload traces are obtained from the platforms with different configurations. For example, NASA iPSC is obtained from the cluster system with each node composed of one CPU; SDSC BLUE is obtained from the cluster system with each node composed of eight CPUs. In our experiments, we scale workload traces with different values to the same configuration of which each node owns one CPU.

The configurations of the experiments are as follows:

1) The scheduling policy: for DawningCloud, SSP and DCS system, we choose the first fit scheduling policy for HTC. The first-fit scheduling algorithm scans all the queued jobs in the order of job arrival and chooses the first job, whose resources requirement can be met by the system, to execute. For MTC workload, firstly we generate the job flow according to the dependency constraints, and then we choose the FCFS (First Come First Served) scheduling policy. The DRP system takes no scheduling policy, since all jobs run immediately without queuing.

2) The time unit of leasing resources: Because when the resource provider assigns or reclaims nodes, it will trigger the setup work, such as wiping off operating system, or deploying software, for cloud systems, we set a quit long time unit: one hour to decrease the management overhead. This factor is same for DawningCloud, the SSP and DRP systems. In fact, EC2 also charge resource with this time unit. The DCS system does not need this factor, since the service provider owns resources.

3) The configurations of the runtime environment in the SSP and DCS systems: since the maximal resource requirements of the NASA and BLUE traces are respectively 128 and 144 nodes, we respectively set the configurations of the runtime environment for NASA and BLUE traces as 128 and 144 nodes. For the Montage workload, because the accumulated resource demand in most of the running time is 166 nodes, we set the configurations of the runtime environment as 166 nodes to improve the throughputs in terms of tasks per second.

4) For DawningCloud, we choose the resource management and provision policies stated in Section 3.2.2. The resource management and provision policies of the SSP and DRP systems are simple. The DRP system depends on each end user's manually requesting and releasing resources. The runtime environment in the SSP system obtains or releases the resources with the fixed size as a whole at the startup and finalization of the RE.

## 4.5 Experiment Results

### 4.5.1 DawningCloud's Parameters Setting
In the DawningCloud, there are two tuning parameters for resource management and provision policies, one is the initial resources, which is represented as B, and the other is the threshold ratio of obtaining resources, which is represented as R.

For HTC workloads, we tune two parameters through changing B from 10 to 80, and R from 1.0 to 2.0. Figure 9 and Figure 10 show the effect of different parameters on two different workload traces. To save the resource consumption and improve the throughputs, we choose *B80_R1.5 as the final configuration for BLUE trace* and *B40_R1.2 as the final configuration for NASA trace.*

For MTC workload, we tune two parameters through changing B from 10 to 80 and R from 2 to 16. Figure 11 shows the effect of different parameters on the Montage workload. To save the resources consumption and improve the throughputs, we choose *B10_R8 as the final configuration for the Montage workload.*

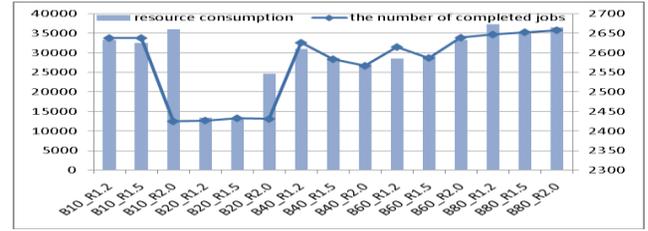

**Figure 9. Resource consumption and the number of completed jobs VS. different parameters setting for BLUE trace. Resource consumption is in term of node*hour.**

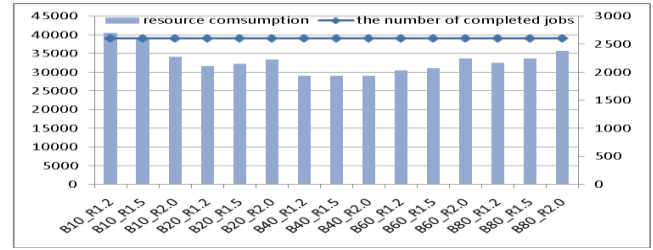

**Figure 10. Resource consumption and the number of completed jobs VS. different parameters setting for NASA trace. Resource consumption is in term of node*hour.**

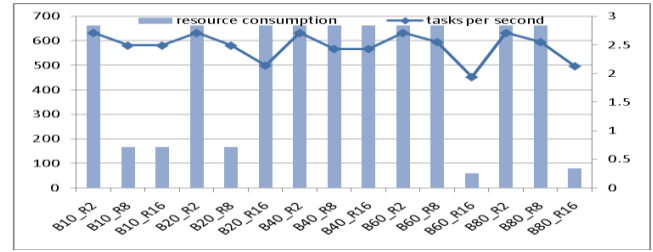

**Figure 11. Resource consumption and tasks per second VS. different parameters setting for Montage workload. Resource consumption is in term of node*hour.**

### 4.5.2 The Evaluation Metric of Service Providers
Table 2, Table 3 and Table 4 summarize the experiment results of two HTC service providers and one MTC service providers with DawningCloud, the SSP, DCS system and DRP systems. The percent of the saved resources are obtained against the resource consumption of the DCS system.

For the DCS and SSP systems, they have the same configurations with the only difference in that the service providers in the DCS system own resources while the service providers in the SSP system lease resources, so they gain the same performance. However, the service providers in these two systems have different total cost of ownership, and we will compare a real case in Section 4.5.5.

For the NASA trace and the BLUE trace, in comparison with the DCS/SSP system, the service providers in the DawningCloud saves the resource consumption maximally by 32.5% and minimally 27.2%, and at the same time gains the same or better throughputs. This is because the service providers in the DawningCloud can resize resources according to the workload status, however the service providers in the DCS/SSP systems owns or leases the resources with the fixed size.

**Table 2. The metrics of the service providers for NASA trace**

| configuration | number of completed jobs | resource consumption | saved resources |
|---|---|---|---|
| DCS system | 2603 | 43008 | / |
| SSP system | 2603 | 43008 | 0 |
| DRP system | 2603 | 54118 | -25.8% |
| DawningCloud | 2603 | 29014 | 32.5% |

**Table 3. The metrics of the service provider for BLUE trace**

| configuration | number of completed jobs | resource consumption | saved resources |
|---|---|---|---|
| DCS system | 2649 | 48384 | / |
| SSP system | 2649 | 48384 | 0 |
| DRP system | 2657 | 35838 | 25.9% |
| DawningCloud | 2653 | 35201 | 27.2% |

**Table 4. The metrics of the service provider for Montage**

| configuration | tasks per second | resource consumption | saved resources |
|---|---|---|---|
| DCS system | 2.49 | 166 | / |
| SSP system | 2.49 | 166 | 0 |
| DRP system | 2.71 | 662 | -298.8% |
| DawningCloud | 2.49 | 166 | 0 |

For the NASA trace and the BLUE trace, in comparison with the DRP systems, DawningCloud can save the resource consumption maximally by 46.4% for the service providers. This is because the dynamic resource negotiation and queuing based resource sharing mechanisms of DawningCloud lead to the decrease of resource consumption. On the other hand, in DRP, each end user directly obtains the resources from the resource provider, which results in that DRP consumes more resources than DawningCloud, but gains the same or better throughputs.

For the NASA trace and the BLUE trace, in our experiments, the DRP system achieves the similar resource consumption as DawningCloud for BLUE workload trace, but consume more resource for NASA workload trace, because that the job arriving frequency of NASA workload trace are smooth among days. The queuing mechanism of DawningCloud can maintain steady resource utilization, which lets DawningCloud save more resources than the DRP system; but the job arriving frequency and system load of BLUE workload trace fluctuate dramatically, which lets the resource utilization of DawningCloud fluctuates too.

For MTC workload, DawningCloud has the same performance of the DCS/SSP system for the service provider. Because that the resource management and provision policies of DawningCloud will dynamically adjust the resources size of the RE according to the accumulated resource demand of jobs in queue, which is same as the chosen configurations of the RE in the DCS/SSP system, as we explained in Section 4.4. After the initial running of Montage trace, DawningCloud adjusts the resources size of the RE to the configurations of the RE in the DCS/SPP system, which results in that DawningCloud has the same performance as that of the DCS/SSP system for the service provider.

For MTC workload, DawningCloud saves the resource consumption by 74.9% in comparison with that of the DRP system for the service provider. This is because the required resources of end users will be provisioned immediately in the DRP system and the peak resource demand of MTC workload is high.

### 4.5.3 The Metric for Resource Provider

Figure 12 and Figure 13 show the experiment results for the resource provider with DawningCloud, the SSP, DRP and DCS systems.

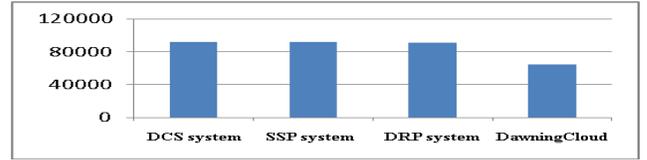

**Figure 12. Total resource consumption of the resource provider. Y-axis is in terms of node*hour.**

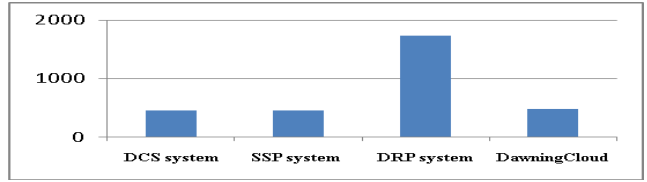

**Figure 13. Peak resource consumption of the resource provider. Y-axis is in term of nodes per hour.**

For the DCS and SSP systems, they have the same performance for the resource provider.

For the resource provider, DawningCloud saves the total resource consumption by 29.7% of that of the DCS/SSP system. In the DCS/SSP system, the service providers lease or purchase resources with the fixed size that is decided according to the peak resource demand of the largest job. In contrast, in DawningCloud, the service providers can start with the small-sized resources and dynamically resize the provisioned resources according to varying resource demand. Hence, the total resource consumption of the resource provider in the DawningCloud is less than that in the DCS/SSP system when workloads of three service providers are consolidated. At the same time, the peak resource consumption of DawningCloud is only 1.06 times of that of DCS/SSP systems.

For the resource provider, DawningCloud saves the total resource consumption by 29.0% of that of the DRP system, and the peak resource consumption of DawningCloud is only 0.21 times of that of the DRP system. Because the required resources will be provisioned immediately in the DRP system, the peak resource consumption of the DRP system is larger than that of DawningCloud.

### 4.5.4 Management Overhead

For the DRP and DawningCloud systems, allocating or reclaiming nodes or VMs will trigger the setup action, e.g. wiping off operating system or data, so it will incur the management overhead for the resource provider. We use the *accumulated times of adjusting nodes that are obtained or released by service providers,* to evaluate the management overhead.

Figure 14 shows the accumulated size of adjusting nodes. We can observe that the SSP system has the lowest management overhead, since it obtains or releases resources only at the startup and the finalization of the RE. DawningCloud has smaller accumulated size of adjusting nodes than that of the DRP system, since the initial resources will not be reclaimed until a runtime environment is destroyed.

In our real test, the total cost of adjusting one node is 15.743 seconds. Excluding wiping off OS, adjusting one node includes the operation of stopping and uninstalling previous RE packets, installing and starting new RE packets. The average overhead of DawningCloud for resource provider is approximately 341 seconds per hour which is acceptable.

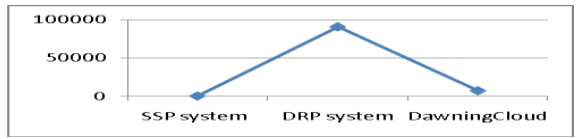

**Figure 14. The accumulated times of adjusting nodes. Y-axis is in terms of times.**

### 4.5.5 Total Cost Ownership of the Service Provider in the SSP and DCS Systems

In this section, we compare the total cost ownership (TCO) of a service provider in the SSP and DCS systems.

For the DCS system, we take a real case from the grid lab of Beijing University of Technology, which is deployed in 2006. The DCS system is composed of 15 nodes, and each node has 2*2 GHZ CPU, 4 GB memory and 160 GB DISKs; the depreciation cycle of system is 8-year; the total capital expenses (*CapEx*) of DCS is 120,000$. Among the operation expenses, the total maintenance cost afforded to the company is 30,000$. The energy and space cost of the DCS is about 1,600$ per month.

For the SSP system, we choose the pricing of Amazon's EC2 Service [3] as the pricing meter. The configuration of one EC2 instance is: 2G CPU, 1.7 GB memory and 140 GB DISK; the price of the EC2 service is 0.1$ per instance * hour and 0.1$ per GB inbound transfer * month.

We calculate the TCO per month of the service provider in the DCS system as follows:

$TCO_{dcs}$ = *(CapEx depreciation) + OpEx (1)*

The TCO of the service provider in the DCS system is *3,160$* per month.

We calculate the TCO per month of the service provider in EC2 as follows:

$TCO_{ssp}$ = *(Total Instance Cost) + (Inbound transfer Cost) (2)*

In order to match the configuration of the DCS system, we choose 30 EC2 instances for the service provider in EC2. The total cost of the instances is: 30day *24hours *30instances *0.1$ =2160$. From the system log, we can know that the average data transfer per month is less than 1000 GB, so the upper cost of inbound transfer is: 1000*0.1=100$. For the SSP system (EC2), the TCO of the service provider is *2,260$* per month, *which is only 71.5% of that of the DCS system*.

### 4.5.6 Analysis

Now we answer the question raised at the beginning of the paper. *Do MTC or HTC service providers benefit from the economies of scale on the cloud platform?*

We have two conclusions: first, from the perspectives of service providers, comparing with the DCS system, SSP is more cost-effective, this is because service providers have the same performance, but the TCO of the service providers in the SSP system is less than that in the DCS system.

Second, with the dynamic resource management mechanism and policies, DawningCloud outperform another two cloud solutions: SSP and DRP from the perspectives of service providers and the resource provider.

Thus, we can conclude: with the enabling system: DawningCloud, MTC or HTC service providers benefit from the economies of scale on the cloud platform

## 5. RELATED WORK

There are two proposed usage models for cloud computing in *MTC* or *HTC* community. Deelman *et al.* [10] propose each staff of an organization to directly lease virtual machine resources from EC2 in a specified period of running applications. Our experiment results show that the system leads to high peak resources consumption, and raises challenge for the capacity planning of system. Evangelinos *et al.* [5] propose that the organization as a whole rents resources with the fixed size from EC2 to create a virtual cluster system that is deployed with the queuing system, like OpenPBS, for HTC workloads. In this model, a service provider as a whole leases the resources with the fixed size from the resource provider, deploys a PBS-like queuing system, and provides job-execution services for end users. Our experiment results show that this system leads to high resource consumption because of its static resource management policy.

Previous efforts fail to propose the enabling system with the autonomic management mechanism to facilitate the resource provider to consolidate MTC and HTC workloads: EC2 [3] directly provides resources to end users, and relies upon end user's manual management of resources; EC2 extended services: RightScale [5] provides automated cloud computing management systems that helps you create and deploy only *Web service applications* that run on EC2 platform; Irwin *et al.* [20] [13] propose a prototype of service oriented architecture for resource providers and consumers to negotiate access to resources over time. However, these previous efforts seldom propose the autonomic management system to consolidate MTC and HTC workloads.

Armbrust *et al.* [2] in theory show the workloads of Web service applications can benefit from the economies of scale of cloud computing system. Our previous work, PhoenixCloud [12] [21], shows the consolidation of Web service applications and parallel batch jobs can decrease the total resource consumption from the perspective of service providers and the resource provider. However, no previous work answers this key question: do MTC or HTC service providers benefit from the economies of scale?

Resource management are widely researched in the context of cloud computing and grid computing. In the context of cloud computing, the work [20] of Duke University designs the Winks scheduler to support a weighted fair sharing model for a virtual cloud computing utility. The goal of the Winks algorithm is to satisfy these requests from a resources pool in a way that preserves the *fairness* across flows; in grid computing, the work [22] proposes the algorithm for scheduling mixed workloads in multi-grid environments, whose goal is to minimize the task's *turnaround time* in grid environment. However, we focus on the resource management for the mix workloads of MTC and HTC, which are not concerned by the previous work.

## 6. CONCLUSION AND FUTURE WORK

In this paper, we have answered two related key questions to the success of cloud computing: for small or medium organizations, can we consolidate their MTC and HTC workloads on a large cloud platform? And on the cloud platform, do MTC or HTC service providers benefit from the economies of scale? Our contributions are three-fold: first, we proposed the dynamic service provision (DSP) model in cloud computing. In the DSP model, the resource provider can create the specific runtime environments on the demand for MTC or HTC service providers, while the service provider can dynamically resize the provisioned resources of the runtime environment. Second, based on the DSP model, we designed and implemented an enabling system, DawningCloud, which provides automatic management for heterogeneous MTC and HTC workloads. Third, our experiments proved that for typical MTC and HTC workloads, MTC and HTC service providers and the resource service provider can benefit from the economies of scale on the cloud platform,.

In the near future, we will focus on building a more formal framework to model and discuss the generalized case in that *n* resource provider provisions resources to *m* service providers of heterogeneous workloads. With the support of this framework, we investigate the optimal resource management and scheduling policies in the context of cloud computing.

## 7. ACKNOWLEDGEMENTS

This paper is supported by the 863 program (Grant No. 2009AA01Z128) and the NSFC programs (Grant No. 60703020 and 60933003).